\documentstyle[12pt]{article}

\begin{document}

\title{{\Large\it 
 Rerum Universitas  Sententia  ex Susy}\thanks{A View of the 
Universe according to {\em Supersymmetry}}
\thanks{(Revided) Report written for the 1996 Awards for Essays on 
Gravitation 
(Gravity Research Foundation)}}
\author{{\large\sf  P.V. Moniz}\thanks{e-mail: {\sf prlvm10@amtp.cam.ac.uk; PMONIZ@Delphi.com}}\\ 
DAMTP,  
University of Cambridge\\ Silver Street, Cambridge, CB3 9EW, UK}
\date{DAMTP -- R96/13}
\maketitle

\vspace{-1cm}

\begin{abstract}

The question if  conserved  currents  can  be sensibly 
defined in supersymmetric minisuperspaces  is investigated in this essay. 
The objective is to employ exclusively the differential equations obtained   
{\em directly} from the Lorentz and supersymmetry quantum 
constraints. 
The 
``square-root'' structure   of 
N=1 supergravity is the motivation 
 to contemplate this  tempting  idea.
However, it is shown that 
such prospect 
 is not feasible but for some very simple scenarios.
 Otherwise, conserved currents (and consistent probability 
densities) can be derived from subsequent Wheeler-DeWitt like equations 
obtained from  the supersymmetric algebra of constraints. 

\end{abstract}

\indent

Philosophers, theologians and scientists have long been 
pondering on     
 the origin, evolution and purpose of our universe \cite{SWH}. 
Einstein's theory of general relativity, quantum mechanics and 
particle physics models represent overwhelming  breakthroughs 
towards providing  answers to those longstanding questions \cite{AA1}-\cite{AA3}. 
A {\em natural} development would be   a 
theory of quantum gravity, which constitutes one of the foremost aspirations 
in theoretical physics. 

The purpose of quantum gravity is to apply the principles of quantum 
mechanics to the {\em entire} universe. Basically, one has to adjoin 
definite laws of initial conditions  
with suitable laws  governing its evolution. Several 
approaches have been provided 
  \cite{H82}--\cite{Vilen84a} 
and their conceivable aim is to allow for  a {\em complete}
explanation of all cosmological observations.

The  inclusion of    supersymmetry may also 
  yeld significant benefits.
Supersymmetry is   a transformation which relates bosons and 
 fermions  
\cite{6,8}. Its  promotion to a gauge symmetry 
 resulted in   an elegant  field theory:  {\rm supergravity} \cite{6,8}.
Supersymmetry may   play an important role when dealing   
divergences in quantum  gravity \cite{9} 
 and removing Planckian masses 
induced by wormholes \cite{HWH,A10}. Furthermore,  
it 
would  be adequate to consider bosons and their 
  fermionic partners  on an equivalent  level
when  studying the very  early universe 
.


N=1 supergravity \cite{6,8} constitutes a  ``square-root''
\cite{2}
of 
 gravity: it   is 
sufficient  to just solve
the Lorentz and supersymmetry constraints
\cite{3,4}.  The 
algebra of constraints implies that a 
physical wave functional $ \Psi $ will consequently  obey 
 the Hamiltonian constraints\footnote{For a review on the 
canonical quantization of supergravity theories see, e.g.,    
ref. \cite{rev}.}. 
The   supersymmetry and Lorentz constraints lead 
  to  differential equations, which are of {\em first-order} 
in the bosonic variables. 
Such  relation temptingly suggests that the possibility to 
derive sensible conserved currents (and positive-definite 
probability densities) should be explored, 
similarly to the procedure intertwining 
 the Klein-Gordon and Dirac equations \cite{Kaku}.

The Wheeler-DeWitt equation  has 
associated with it a conserved current \cite{G26a}
$
J \sim \Psi^* \nabla \Psi - \Psi \nabla \Psi^*
$. 
It satisfies 
$
\nabla J = 0,
$
where $\nabla $ is the Laplacian in superspace \cite{G26a}. 
A conserved probability 
can be defined from $J$ 
but it  can  be afflicted   with  
negative values  \cite{AA2,Vilen82,G26a,wh7}.
A possible solution   requires a suitable choice of 
hypersurfaces.
Such procedure   works {\it only}  within 
a  semiclassical minisuperspace approximation\footnote{This is probably sufficient 
for all {\em practical}
 purposes. 
In the semiclassical case, the wave function is of the WBK form $\Psi \sim 
C e^{-I}$, where $I$ and $C$ are both complex, $I = I_R - iS$ and $|\nabla S| \gg 
|\nabla I_R|$. I.e.,   $S$ will be an {\it approximate} solution of the 
Lorentzian Hamilton-Jacobi equation. In particular,  
$\nabla J =0$ is obtained with 
$
J \sim e^{-I_{R}} |C^2| \nabla S
$ \cite{AA2,G26a}.} \cite{Vilen82,wh7,G26c}.

A closed  FRW    supersymmetric minisuperspace 
model with a scalar multiplet  will be used  
in this essay. 
Several improvements concerning previous results in 
ref. \cite{A8,A22,A17} are included as well. 
The action of the more general theory of N=1 supergravity in the 
presence of gauged supermatter 
(see eq. (25.12) in ref. \cite{8}) will be employed.
The metric variables are represented  by  
the tetrad: $
 e_{a\mu} = {\rm diag}  \left(
 N (\tau),  a E_{\hat a I} \right)$. 
Here $ \hat a $ and $ i $ run from 1 to 3 and 
$ E_{\hat a i} $ is a basis of left-invariant 1-forms on the unit $ S^3 $
with volume $ \sigma^2 = 2 \pi^2 $.
This   reduces the number of degrees of freedom provided by $ e_{AA'
\mu} $, 
the spinorial form of the 
tetrad. A 
suitable choice for the gravitino fields, 
 $ \psi^A_{~~\mu} $ and $ \bar\psi^{A'}_{~~\mu}$, 
is then required. 
Hence $\psi^A_{~~0} $ and $ \bar\psi^{A'}_{~~0} $ will  be functions of time only 
and $
\psi^A_{~~i} = e^{AA'}_{~~~~i} \bar\psi_{A'}(t) ~, ~
\bar\psi^{A'}_{~~i} = e^{AA'}_{~~~~i} \psi_A (t)$,  
where  the new spinors $ \psi_A $ and their 
Hermitian  conjugate $\bar\psi_{A'} $  are introduced 
\cite{rev,A8, A22,A18a}.  
The scalar supermultiplet, consisting of a complex  scalar 
field $ \phi, \bar\phi $ and spin-$\frac{1}{2}$ fields $ \chi_A, \bar\chi_{A'}$ 
are chosen to   depend only on time. The  
remaining fields  are taken henceforth to be zero.

The analysis becomes simpler if  the  fermionic fields, $ \chi_{A} $,  
$ \psi_{A} $ are redefined as follows:  
$  
 \hat \chi_{A} = {\sigma a^{3 \over 2} \over 2^{1 \over 4} (1 + \phi \bar \phi)} \chi_{A}~, ~  
  \hat \psi_{A} = {\sqrt{3} \over 2^{1 \over 4}} \sigma a^{3 \over 2} \psi_{A}$, ~
 and similarly for their   Hermitian conjugates. 
In addition,  
 unprimed spinors will be used: 
$ \bar \psi_{A} = 2 n_{A}^{~B'} \bar \psi_{B'}~, ~
 \bar \chi_{A} = 2 n_{A}^{~B'} \bar \chi_{B'}$. 
The 
 coordinates of the configuration space are chosen to be 
$ (\chi_{A} , \psi_{A} , a , \phi , \bar \phi) $  and 
$ (\bar \chi_{A} , \bar \psi_{A}$, 
  $\pi_{a}$ , $\pi_{\phi}$ , $\pi_{\bar \phi}) $ form 
 the momentum operators in this representation.
Quantum mechanically  (with $\hbar =1$):  
\begin{equation}  
 \bar \chi^{A}   \rightarrow  
-{  \partial   \over   \partial   \chi^{A}} ,~ \bar \psi_{A} \rightarrow {  \partial   \over   \partial   \psi^{A}} , 
     \pi_{a}  \rightarrow 
 {  \partial   \over   \partial   a} , ~\pi_{\phi} \rightarrow -i {  \partial   \over   \partial   \phi},~
 \pi_{\bar \phi} \rightarrow -i {  \partial   \over   \partial   \phi} ~.
\label{eq:2.12}
\end{equation}
The Lorentz 
constraint $ J_{AB} = 0 $ 
takes the form 
  \begin{equation} 
 J_{AB} = \psi_{(A} \bar{\psi}_{B)} - \chi_{(A} \bar{\chi}_{B)}
 =  0~. 
\label{eq:2.14}
\end{equation}
This constraint implies that  the 
 most general form for the wave function 
of the universe is 

\begin{equation} 
  \Psi = A  +  B \psi^{C} \psi_{C} + C \psi^{C} \chi_{C} + 
D \chi^{C} \chi_{C} + E \psi^{C} \psi_{C} \chi^{D} \chi_{D} 
\label{eq:2.15}
\end{equation}
where $A$, $B$, $C$, $D$, $E$   are functions of $a$, $\phi$ ,$\bar \phi$ 
  only.

The following eight equations result from the 
application of the supersymmetry constraints $S_A, \bar S_{A'}$ 
on $\Psi$ given by (\ref{eq:2.15}) (see ref. \cite{A22}):

\begin{eqnarray}  
  { a \over 2 \sqrt{6}} {  \partial   A \over   \partial   a} + \sqrt{3 \over 2} \sigma^{2} a^{2} A =0~,~
  {a \over \sqrt{6}} {   \partial   E \over   \partial   a}  - 
 \sqrt{6} \sigma^{2} a^{2} E & = &  0~, \label{eq:2.2.4} \\
\frac{\partial A}{\partial r}   -  
 i\frac{1}{r}\frac{\partial A}{\partial \theta} = 0 ~,~
\frac{\partial E}{\partial r} +   
 i\frac{1}{r}\frac{\partial E}{\partial \theta} &=& 0~. \label{eq:2.2.20e}
\end{eqnarray}
\begin{eqnarray}
   (1 + \phi \bar \phi) {   \partial   B \over   \partial   \phi}  + 
  {1 \over 2} \bar \phi B 
+ { a \over 4 \sqrt{3}} {  \partial   C \over   \partial   a} - {7 \over 4 \sqrt{3}} C + 
{\sqrt{3} \over 2}  \sigma^{2} a^{2} C & = &  0~, \label{eq:2.2.5} \\
{a \over \sqrt{3}} {  \partial   B \over   \partial   a}  - 
 2 \sqrt{3} \sigma^{2} a^{2} B - \sqrt{3} B
 + (1 + \phi \bar \phi) {  \partial   C \over   \partial   \bar \phi} + {3 \over 2} \phi C 
&=& 0 
~,\label{eq:2.2.6} \\ 
{a \over \sqrt{3}} {  \partial   D \over   \partial   a} 
+   2 \sqrt{3} \sigma^{2} a^{2} D - \sqrt{3} D 
- (1 + \phi \bar \phi) { \partial C \over \partial \phi} - {3 \over 2} \bar \phi C & =  &
0~, \label{eq:2.2.7} \\ 
(1 + \phi \bar \phi) {  \partial   D \over   \partial   \bar \phi}  + 
  {1 \over 2} \phi D 
- {a \over 4 \sqrt{3}} {  \partial   C \over   \partial   a} + {7 \over 4 \sqrt{3}} C + {\sqrt{3} \over 2} \sigma^{2} a^{2} C & = &  0 ~. 
\label{eq:2.2.8}
\end{eqnarray}
Notice that $r^2 \equiv \phi\bar\phi$  with $\phi \equiv r e^{i\theta}$ 
was employed\footnote{Using $\phi, \bar\phi$ does {\it not} allow to find the 
explicit dependence of $\Psi$ (see ref. \cite{rev,A22,A23,A17}). 
An alternative approach 
is to 
write 
$\phi = r e^{i\theta}$ and hence to effectively decouple the 
two degrees of freedom  associated with the complex 
scalar field.
It should be 
stressed that this procedure  has  {\it not} yet 
been employed {\it directly} in the 
supersymmetry constraints but rather on the 
Hamiltonian constraints \cite{A10,Khala}.} in   
eq. (\ref{eq:2.2.20e}).

Eq.   
(\ref{eq:2.2.4}), (\ref{eq:2.2.20e})  
constitute  decoupled equations for $A$ and $E$. 
 Eq. (\ref{eq:2.2.5}) and 
(\ref{eq:2.2.6}) constitute 
  coupled equations between $B$ and $C$, while  eq. (\ref{eq:2.2.7}), 
(\ref{eq:2.2.8})  
are coupled equations between $C$ and $D$.
It can be shown that  
eq. (\ref{eq:2.2.5})-(\ref{eq:2.2.8}) 
imply 
 $ C = 0$ \cite{A22,A17,A23}. 
A two-dimensional spherically symmetric K\"ahler geometry 
\cite{8} has been chosen here but this result seems 
independent of that choice \cite{A22}. 

The following steps are now followed. Multiply the  
 first equation in (\ref{eq:2.2.4}) by $E$, 
and  the second by $A$.
Then add them. Now 
multiply 
  the first equation in 
(\ref{eq:2.2.20e}) by $E$, then the second by $A$ and subtract them. 
Employing now 
  $C=0$,    
 multiply  eq. (\ref{eq:2.2.6})  by $D$ and eq. 
(\ref{eq:2.2.7}) 
 by $B$. Then add them. 
Finally, multiply eq. (\ref{eq:2.2.5}) by $D$, eq. (\ref{eq:2.2.8}) by 
$B$ and subtract them. The overall result is written as 
\begin{eqnarray}
\frac{\partial (A\cdot E)}{\partial a} + 
\frac{\partial (A\cdot E)}{\partial \theta} - ir  \left(
\frac{\partial E}{\partial r} A - \frac{\partial A}{\partial r} E\right) & = &  0,
\label{eq:2.New22a} \\
D_a (B\cdot D)  + 
\frac{\partial (\, B \cdot D)}{\partial \, \theta } 
-ir  \left(\frac{\partial \, B}{\partial \, r} D 
- \frac{\partial \, D}{\partial \, r} B \right)
   & = &   0~,
\label{eq:2.New22b}
\end{eqnarray}
with the generalized derivative $D_a \equiv \partial_a - \frac{6}{a}$.

From eq. (\ref{eq:2.2.4}), (\ref{eq:2.2.20e}) and 
eq. (\ref{eq:2.2.5})--(\ref{eq:2.2.8}) with $\phi = r e^{i\theta}$ 
the quantum state corresponding to a $k=1$ FRW supersymmetric 
model with scalar supermultiplets is given by 
\begin{eqnarray}
\Psi & = &  c_1 r^{\lambda_1} e^{-i\lambda_1 \theta} 
e^{- 3\sigma^2 a^2}  +  
c_3 a^3 r^{\lambda_3} e^{-i\lambda_3 \theta}(1 + r^2)^{\frac{1}{2}} 
e^{ 3\sigma^2 a^2}
  \psi^{C} \psi_{C} \nonumber \\ & + &  
c_4 a^3 r^{\lambda_4} e^{i\lambda_4 \theta}(1 + r^2)^{\frac{1}{2}} 
e^{- 3\sigma^2 a^2} 
\chi^{C} \chi_{C} + 
c_2 r^{\lambda_2} e^{i\lambda_2 \theta}
e^{ 3\sigma^2 a^2}
 \psi^{C} \psi_{C} \chi^{D} \chi_{D} ~,
\label{eq:2.New1}
\end{eqnarray}
where 
  $\lambda_1$...$\lambda_4$  
and $c_1$...$c_4$ are constants. 
 The exponential factors $ e^{\pm 3\sigma^2 a^2}$  in (\ref{eq:2.New1}) 
are to be viewed as   $e^{\pm I}$, where $I$ is the  Euclidean action 
for a classical solution {\em without} matter outside or inside a 
three-sphere with radius $a$ (see ref. \cite{A8}). In the absence of matter,  
the Hartle-Hawking state \cite{HH83} for this model is therefore 
given by $\Psi_{HH} = \psi_A\psi^A e^{3\sigma^2 a^2}$.  
 A solution $\Psi_{WH} = e^{-3 \sigma^2 a^2}$ bears quantum 
wormhole properties \cite{A10,13,A23}.  
However,  the full  physical interpretation of the 
bosonic coefficients in (\ref{eq:2.New1})  is less clear. 
$C=0$ seems to imply that while a fermionic 
state $\chi^A\psi_A$ is allowed by Lorentz invariance, 
supersymmetry effectively rules it out since  the spin-$\frac{1}{2}$ 
$\chi^A$, $\psi_A$ fields have different roles. 
In addition,  
the scalar field dependence is {\em different} from the expressions in 
non-supersymmetric  quantum 
FRW models  with complex scalar fields (cf. ref. \cite{Khala}).  

The presence of the last term in both eq. (\ref{eq:2.New22a}), 
(\ref{eq:2.New22b}) clearly prevent us to associate them with 
an equation of the type of $\nabla J = 0$. Notice that 
eq. (\ref{eq:2.2.20e}) and (\ref{eq:2.2.5}), (\ref{eq:2.2.8}) 
lead directly to $\frac{\partial (A\cdot E)}{\partial \theta} - ir  \left(
\frac{\partial E}{\partial r} A - \frac{\partial A}{\partial r} E\right)  =  0  $ and 
$\frac{\partial (\, B \cdot D)}{\partial \, \theta } 
-ir  \left(\frac{\partial \, B}{\partial \, r} D 
- \frac{\partial \, D}{\partial \, r} B \right)
   = 0 $, respectively. 
This feature is quite relevant. In contrast with 
the case of a FRW model with complex scalar fields in 
non-supersymmetric quantum cosmology \cite{Khala}, 
the variable $\theta$ is no longer a cyclical 
coordinate when supersymmetry is present. 

In usual quantum cosmology with complex scalar fields,   
the corresponding action for FRW models is invariant under a 
$\phi \mapsto  e^{i\alpha} \phi$ transformation.  Hence,  the 
conjugate  momentum  $\pi_\theta \sim ir^2 \frac{\partial \, \theta}{\partial \, t }$
is a constant  \cite{Khala} and $\theta$ is a cyclical coordinate. 
However, in the corresponding 
supersymmetric scenario \cite{A22} there are  terms in the 
action (25.12) of \cite{8} that do {\it not} allow for this 
invariance to be present.  Let us  write the conjugate momenta to $r$ and $\theta$
\begin{eqnarray}
\pi_r &= & 2 \frac{\partial \, r}{\partial \, t} \frac{\sigma^2 a^3}{(1 + 
r^2)^2} - \frac{\sigma^2 a^3 e^{-i\theta}}{\sqrt{2}(1 + r^2)^2} 
\left(\chi^A\psi_{0A} + 3 n_{AA'} \chi^A \bar\psi^{A'} \right) 
\nonumber \\
& - & 
\frac{\sigma^2 a^3 e^{i\theta}}{\sqrt{2}(1 + r^2)^2} 
\left(\bar\chi_{A'}\bar\psi_{0}^{A'} - 3 n_{AA'} 
\bar\chi^{A'} \psi^{A} \right)~,
\label{eq:2.2.pir}
\end{eqnarray}
\begin{eqnarray}
\pi_\theta & = & 2\frac{\partial \, \theta}{\partial \, t} 
\frac{\sigma^2 r^2 a^3}{(1 + r^2)^2} + 
\frac{5 \sigma^2 r^2 a^3}{\sqrt{2}(1 + r^2)^3}
n^{AA'} \bar\chi_{A'}\chi_A \nonumber \\
& - &  \frac{3 \sigma^2 r^2 a^3}{\sqrt{2}(1 + r^2)}
n^{AA'} \psi_A \bar\psi_{A'}\nonumber \\
& + & \frac{i r \sigma^2 a^3 e^{-i\theta}}{\sqrt{2}(1 + r^2)^2}
\left(\chi^A\psi_{0A} + 3 n_{AA'}\chi^A\bar\psi^{A'}\right)
\nonumber \\
& - & \frac{i r \sigma^2 a^3 e^{i\theta}}{\sqrt{2}(1 + r^2)^2}
\left(\bar\chi^{A'}\bar\psi_{0}^{A'} - 3 n_{AA'}
\bar\chi^{A'}\psi^{A}\right)~,
\label{eq:2.2pitheta}
\end{eqnarray}
and notice that the usual procedure ${\cal H} \sim p\dot{q} - L$ involves  
a term like $\sigma^2 a^3 \left[ \left(\frac{\partial \, r}{\partial \, t}\right)^2 + ir^2 
\left(\frac{\partial \, \theta}{\partial \, t}\right)^2 \right]$. 
It is precisely the last two  terms in (\ref{eq:2.2.pir}) and 
(\ref{eq:2.2pitheta}) that allow   the supersymmetry constraints to 
be obtained explicitely 
from the coefficients in $\psi^A_0,\bar\psi^{A'}_0$ in the 
Hamiltonian $\cal H$.
But the last two terms in both 
(\ref{eq:2.2.pir}) and (\ref{eq:2.2pitheta}) are also a 
direct 
consequence of  the {\it non-invariant } terms in the 
  action \cite{8}. 
Moreover, eq. (\ref{eq:2.2pitheta}) is basically 
translated into the last two terms in eq. 
(\ref{eq:2.New22a}), 
(\ref{eq:2.New22b}). Hence, 
$\theta$ no longer being a cyclical coordinate 
implies that a relation as $\nabla J=0$
{\em cannot} be sensibly defined. Furthermore, 
this fact is inherited from local supersymmetry 
being now a feature of the reduced model. 
Thus, it seems then that there is close relation 
between the absence  of cyclical coordinates, 
conserved currents from $\Psi$ and the presence of 
supersymmetry. 

{\em No}  conserved currents are possible 
to obtain in 
our supersymmetric minisuperspace 
{\em directly}
 from the Lorentz and supersymmetry constraints. 
A relation as $\nabla J = 0$ can only be achieved for the 
simple case of pure N=1 supergravity, where eq. 
(\ref{eq:2.2.4})--(\ref{eq:2.2.8}) are reduced to just 
(\ref{eq:2.2.4}). Consequently, we obtain 
$\frac{\partial (A\cdot B)}{\partial a} = 0$ 
as expected.

Overall, the final message in this essay is the following. 
 Conserved currents   do not 
seem possible to obtain {\em directly} from the supersymmetry 
constraints equations. Only for very simple scenarios does this becomes 
possible. Our results should then be compared with the 
assertions present in ref. \cite{OO,mallet}. 
A Wheeler-DeWitt--like equation becomes mandatory 
when supersymmetry implies a mixing between 
 the  fermionic sectors\footnote{This occurs when 
either Bianchi class-A models or  the full theory of N=1 
supergravity are used \cite{melast}.} in $\Psi$.  
. Only 
then can  
explicit expressions for the bosonic coefficients in $\Psi$ be obtained. 
As a consequence,    the standard steps present in  
usual quantum cosmology ought to be adapted 
\cite{AA2,G26a}. It should also be stressed 
that the wave functional
$\Psi(e^{AA'}_\mu; \psi^A_\mu;...)$ 
for  N=1 quantum supergravity
is a Grassman-algebra-valued 
functional and thus quite different from relativistic quantum 
mechanics wave functions.

\vspace{1cm}

{\large\bf Acknowledgements}

\vspace{0.3cm}

This essay was supported by   a JNICT/PRAXIS--XXI Fellowship BPD/6095/95. 
The author 
is grateful to A.D.Y. Cheng, A. Yu. Kamenshchik and C. Kiefer 
 for useful  comments and discussions. Conversations 
with    R. Graham, H. Luckock and O. Obregon 
also provided motivation and insight concerning 
some of the issues hereby discussed.

\end{document}